# Supercurrent-controlled magnon transport via suppression of the superconducting gap

Tetsuma Mandokoro[1], Yoichi Shiota[1,2*], Hisakazu Matsuki[1,2], Ryusuke Hisatomi[1,2],
Shutaro Karube[1,2], Akito Daido[3], Youichi Yanase[3], and Teruo Ono[1,2,4†]
[1]*Institute for Chemical Research, Kyoto University, Uji, Kyoto 611-0011, Japan*
[2]*Center for Spintronics Research Network, Kyoto University, Uji, Kyoto 611-0011, Japan*
[3]*Department of Physics, Graduate School of Science, Kyoto University, Kitashirakawa, Kyoto 606-8502, Japan*
[4]*International Center for Synchrotron Radiation Innovation Smart, Tohoku University, Sendai, Miyagi 980-8572, Japan*

Understanding interactions between different low-energy excitations is a central theme in condensed-matter physics. Magnons and superconducting condensates represent two distinct macroscopic quantum states, yet their mutual interaction remains largely unexplored. Here, we demonstrate supercurrent-controlled magnon transport in a ferromagnet/superconductor heterostructure. We observe that the magnon transport signal decreases as the supercurrent increases. This suppression originates from a supercurrent-induced reduction of the superconducting gap, which enhances magnon dissipation into the superconductor via Bogoliubov quasiparticles. Our results provide direct experimental evidence of a supercurrent-controlled interaction between magnons and quasiparticles in the superconductor. The approach used in this study relies on supercurrents, enabling magnon modulation without Joule heating. These findings establish a new route to control magnons via superconductivity and pave the way for energy-efficient magnonic devices.

## I. INTRODUCTION

Magnons, the quanta of spin waves in magnetic materials, provide a powerful platform for studying spin transport without accompanying charge motion. Their ability to carry angular momentum over long distances has stimulated extensive research in magnonics, where magnons are explored as potential information carriers for next-generation spintronic technologies [1-5]. A key requirement for magnon-based functionalities is the ability to efficiently control magnon transport. Active control of magnon transport can be achieved by current passing through a ferromagnetic or heavy-metal wire along the magnon propagation path [6-9]. However, such approaches inevitably generate Joule heating, leading to additional energy dissipation. Minimizing such dissipation therefore remains a major challenge in magnonics. One promising strategy is to replace normal-metal control elements with superconductors, which can carry electric current without resistive losses. Although superconductors were traditionally considered incompatible with magnetic materials because of their Meissner effect, recent advances in spintronics have revealed rich interactions between magnetism and superconductivity [10-13]. In particular, spin-triplet Cooper pairs that enable spin transport can be generated at interfaces between superconductors and magnetic materials with non-uniform magnetic textures [14-17]. Furthermore, theoretical studies have predicted that dynamically generated non-uniform magnetic structures, such as spin waves, can also induce spin-triplet Cooper pairs [18,19]. These developments have stimulated increasing interest in the interactions between spin waves and superconductors, raising the possibility of controlling magnon transport via superconducting correlations.

Recent studies have demonstrated that magnons in ferromagnets can excite Bogoliubov quasiparticles in an adjacent superconductor [20,21]. These quasiparticle excitations are governed by the relation between the magnon energy and the superconducting gap. Consequently, the injection of magnons into the superconductor depends on the relative magnitude of the superconducting gap and the magnon energy. This coupling suggests a possible route for controlling magnon transport by tuning the superconducting state. Notably, a supercurrent is known to suppress the superconducting gap [22,23]. Therefore, applying a supercurrent provides a means to continuously modulate the superconducting gap and, consequently, the dissipation of magnons into the superconductor.

In this study, we demonstrate supercurrent-induced modulation of magnon transport in a ferromagnet/superconductor heterostructure. By suppressing the superconducting gap with the applied supercurrent, magnon dissipation into the superconductor is enhanced, resulting in a controlled reduction of the magnon transport. Our results provide experimental evidence of an interaction between magnons and the superconducting state mediated by a supercurrent. Furthermore, by using a model that accounts for changes in the superconducting gap due to the supercurrent, we quantitatively reproduce the experimental results. Since the supercurrent does not generate Joule heating, this approach enables modulation of magnon transport without additional

*shiota-y@scl.kyoto-u.ac.jp
†ono@scl.kyoto-u.ac.jp

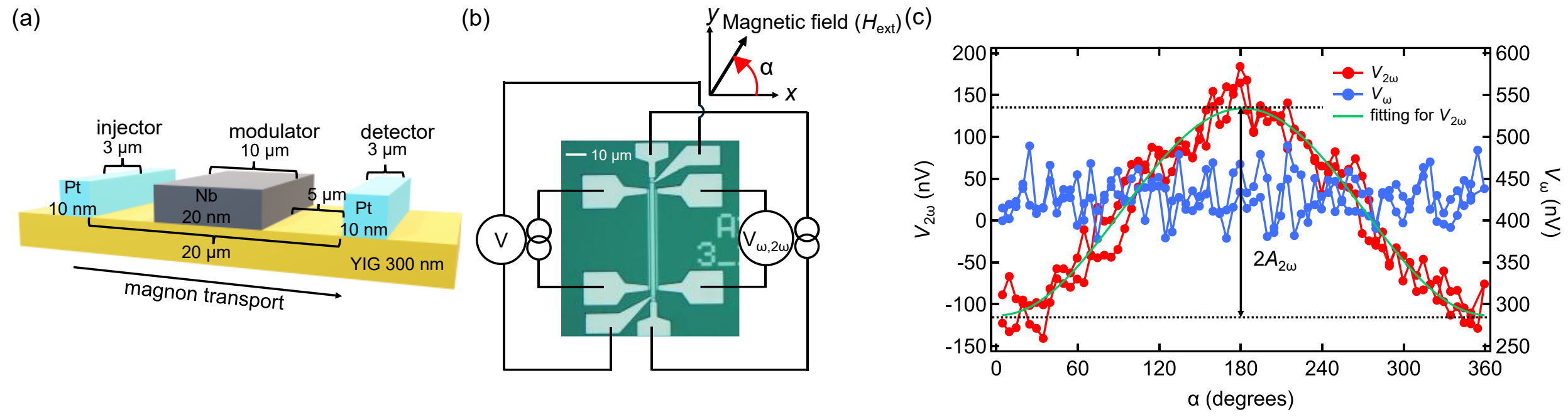

FIG. 1. (a) Schematic illustration of the device. Two Pt electrodes serve as the magnon injector and detector, while an Nb electrode serves as a superconducting modulator. (b) Optical micrograph of the device and measurement configuration. The in-plane magnetic field angle α is defined with the $x$-axis corresponding to $\alpha = 0°$. Current sources are connected to the Pt injector and Nb modulator, and a lock-in amplifier is connected to the Pt detector. A nanovoltmeter is also connected to the Nb electrode to monitor its superconducting state. (c) Magnon transport signals in the YIG/Nb sample at 2 K under $\mu_0 H_{\text{ext}} = 30$ mT as a function of α. Red symbols represent the thermally excited signal ($2\omega$ component), while blue symbols represent the electrically excited signal ($1\omega$ component). The green solid line is the fitting result for the $2\omega$ component.

energy losses, paving the way for the realization of lower-energy magnonic devices.

## II. DEVICE FABRICATION AND MEASUREMENT METHODS

300-nm-thick yttrium iron garnet (YIG) (111) thin films were deposited on a gadolinium gallium garnet (GGG) (111) substrate by radio-frequency (rf) magnetron sputtering. The as-deposited YIG films were subsequently epitaxially crystallized by annealing at 700 °C for one hour in air. Details of the ferromagnetic resonance measurements and the resulting magnetic damping constant are summarized in Appendix A. Device fabrication was then carried out on these YIG films. Two 10-nm-thick Pt electrodes and one 20-nm-thick Nb electrode were fabricated on the YIG by rf magnetron sputtering and a lift-off process to form the injector, detector, and modulator contacts, respectively, as shown in Fig. 1(a). The injector and detector have widths of 3 μm and lengths of 300 μm, respectively, while the modulator has a width of 10 μm and a length of 400 μm. The edge-to-edge distance between the injector and the detector is 20 μm, and that between the modulator and the adjacent electrodes is 5 μm. Prior to these electrodes' depositions, oxygen plasma ashing was performed to remove residual resist from the surface. Contact electrodes of Ti(5 nm) / Cu(100 nm) / Au(20 nm) were then patterned using photolithography and a lift-off process. In addition, to eliminate interactions between the superconductor and magnons at the interface, a control sample was prepared by inserting a 20-nm-thick $AlO_x$ layer between the YIG and Nb layers (Hereafter, the sample without $AlO_x$ and the sample with an inserted $AlO_x$ layer are referred to as the YIG/Nb sample and the YIG/$AlO_x$/Nb sample, respectively). All other fabrication conditions and device geometries were kept identical to those of the YIG/Nb sample. To minimize possible interfacial disorder, Nb and $AlO_x$/Nb were deposited at room temperature.

In the following experiments, the applied magnetic field was set larger than the coercive field of the YIG (approximately 0.2 mT at 2-5 K as detailed in Appendix B) to align the magnetization with the external magnetic field. Figure 1(c) shows the magnon transport signal of the YIG/Nb sample at 2 K. The measurements were performed using the circuit configuration shown in Fig. 1(b). The in-plane angular dependence (α) was measured by rotating the external magnetic field of $\mu_0 H_{\text{ext}} = 30$ mT. An alternating current with a frequency of 17 Hz and an amplitude of 1.2 mA ($4.0 \times 10^{10}$ A/m$^2$) was applied to the injector to excite magnons, and the voltage signal at the detector was measured using a lock-in amplifier. No angular dependence was observed in the first-harmonic ($1\omega$) component (blue symbols), originating from electrically excited magnons. In contrast, a clear angular dependence was observed in the second-harmonic ($2\omega$) component (red symbols), originating from thermally excited magnons. Here, the temperature gradient is generated in the out-of-plane direction by Joule heating in the Pt injector. The observation that the angular dependence appears only in the $2\omega$ component is consistent with previous studies and is attributed to a reduction of the spin mixing conductance at the YIG/Pt interface at low temperatures [24,25]. As a result, only the thermal magnon excitation is observed in the magnon transport signals. The amplitude of the thermal magnon transport signal is evaluated by fitting the data with $-A_{2\omega}\cos\alpha + V_{offset}$ [24], where α is the angle of the external magnetic field shown in Fig. 1(b), and $V_{offset}$ is an offset voltage. The corresponding fitting result is shown as a green solid line in Fig. 1(c).

*shiota-y@scl.kyoto-u.ac.jp
†ono@scl.kyoto-u.ac.jp

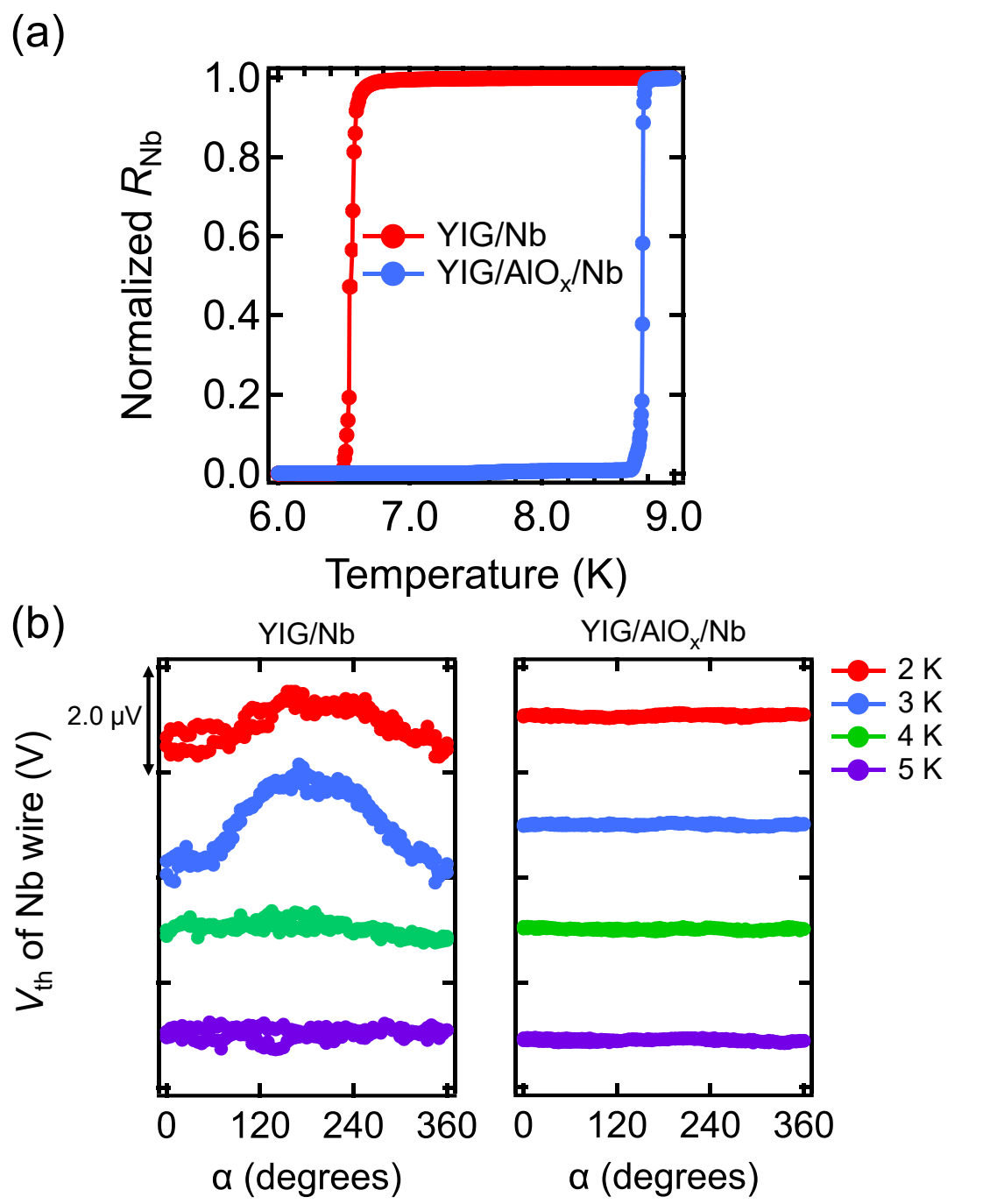

FIG. 2. (a) Temperature dependence of the normalized Nb resistance ($R_{Nb}$) for the YIG/Nb and the YIG/AlO$_x$/Nb sample. (b) Angular dependence of the thermally driven voltage $V_{th}$ of Nb wire under the magnetic field of $\mu_0 H_{ext} = 5$ mT. The curves are vertically offset for clarity.

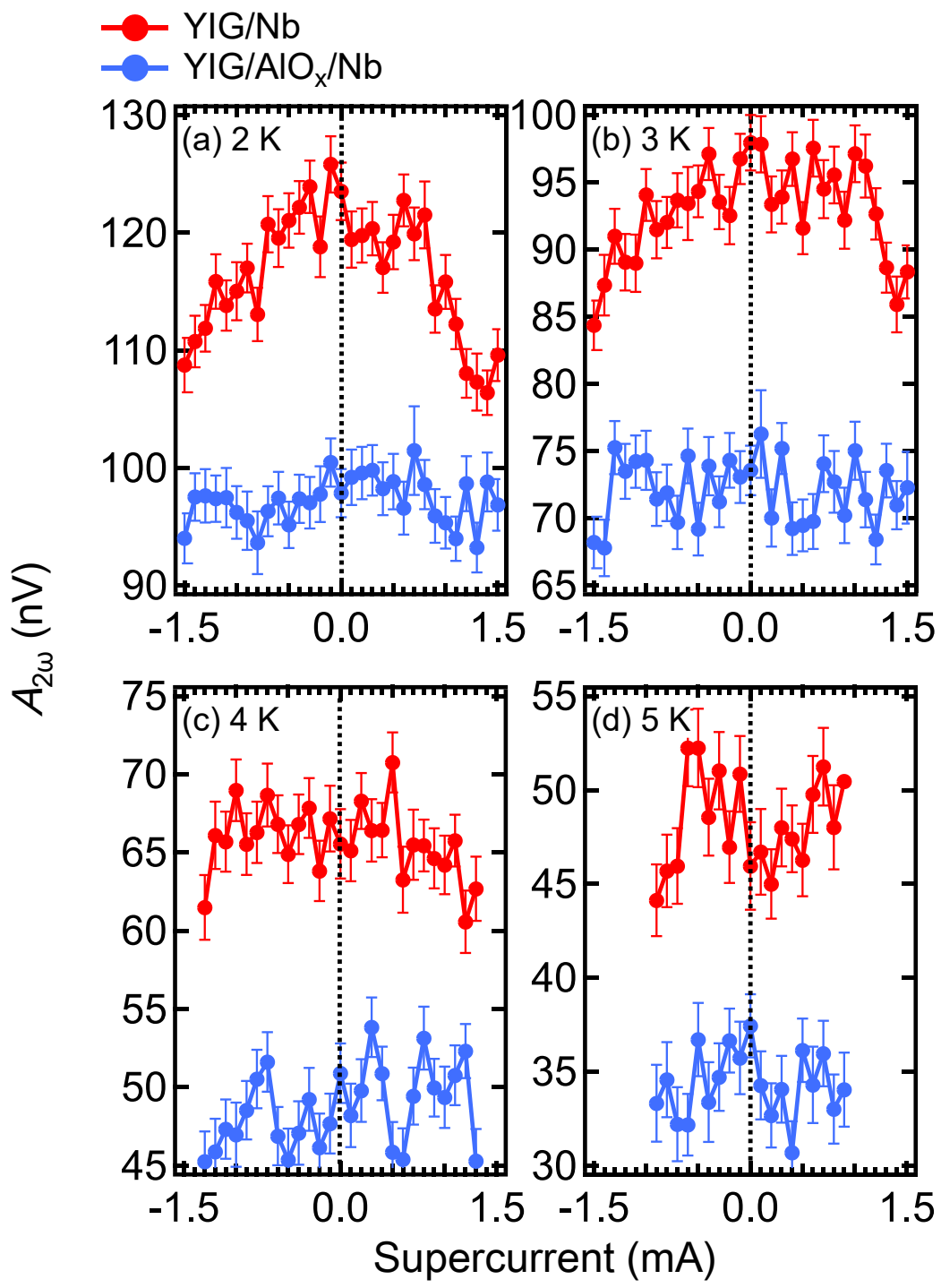

FIG. 3. Supercurrent dependence of the magnon amplitude, $A_{2\omega}$, under the magnetic field of $\mu_0 H_{ext} = 30$ mT. (a)-(d) Comparison of the YIG/Nb sample (red symbols) and the YIG/AlO$_x$/Nb sample (blue symbols) at 2-5 K. Only data acquired while the Nb modulator remains in the superconducting state are shown.

## III. SUPERCONDUCTING PROPERTIES

Next, we characterize the superconducting properties of the Nb wire. Figure 2(a) shows the temperature dependence of the Nb resistance for the YIG/Nb and the YIG/AlO$_x$/Nb samples. The superconducting transition temperature ($T_c$) of the YIG/AlO$_x$/Nb sample is consistent with literature values for Nb thin films [26,27]. However, the $T_c$ of the YIG/Nb sample is approximately 2 K lower than that of the YIG/AlO$_x$/Nb sample. This reduction of $T_c$ in the YIG/Nb sample may suggest the presence of an interfacial exchange field acting on the Nb wire [28], although possible differences in interfacial chemical disorder, such as intermixing or defect formation, may also contribute to the reduction in $T_c$. Figure 2(b) shows the angular dependence of the thermally driven voltage $V_{th} = \frac{V(+I)+V(-I)}{2}$ of the Nb wire, measured when magnons are excited by a direct current $I = 1$ mA applied to the Pt injector under $\mu_0 H_{ext} = 5$ mT, while no current is applied to the Nb wire. Here, $V(I)$ denotes the voltage of the Nb wire measured when a direct current $I$ is applied to the Pt injector. Sinusoidal curves were observed only in the YIG/Nb sample, which originates from the inverse spin Hall effect of quasiparticles in the superconductor induced by magnons [20,21]. The quasiparticle signal is most pronounced at 3 K, whereas the signal at 4 K and 5 K becomes negligible compared with that observed at 2 K and 3 K. In contrast to the YIG/Nb sample, no angle dependence was observed in the YIG/AlO$_x$/Nb sample at any temperature. These results indicate that the quasiparticle excitation originates from magnons and is governed by the interplay between the superconducting gap Δ and the magnon chemical potential $\delta\mu$, consistent with previous studies [20,21].

## IV. SUPERCURRENT-CONTROLLED MAGNON TRANSPORT

*shiota-y@scl.kyoto-u.ac.jp
†ono@scl.kyoto-u.ac.jp

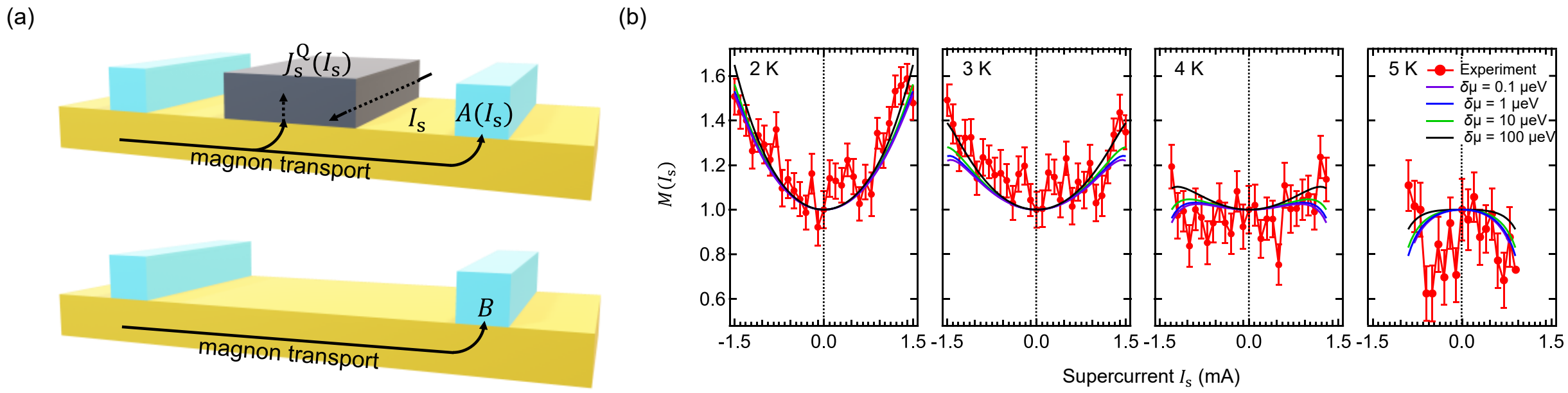

FIG. 4. (a) Schematic illustration of magnon transport and its decomposition. Magnons propagate from the injector to the detector, giving rise to the observed signal $A(I_s)$. In the absence of this dissipation channel, Nb, the total magnon signal is given by $B$. The dissipated component is converted into a quasiparticle spin current in the superconductor, expressed as $J_s^Q \propto B - A(I_s)$. (b) Normalized modulation amplitude $M(I_s)$ as a function of supercurrent. Symbols represent experimental data $\frac{B-A(I_s)}{B-A(0)}$, and solid lines show theoretical predictions $\frac{J_s^Q(I_s)}{J_s^Q(0)}$ calculated from Eq. (3). The solid lines in purple, blue, green, and black show the calculation results for magnon chemical potential $\delta\mu$ = 0.1, 1, 10, and 100 μeV, respectively.

To examine whether magnon transport can be modulated by a supercurrent, we measured the thermal magnon transport signal while applying a current to the Nb modulator. Figure 3 shows the amplitude of the magnon transport signals, $A_{2\omega}$ as a function of the applied supercurrent in the Nb modulator. The applied current corresponds to a current density of $5.0 \times 10^9$ A/m$^2$ per 1 mA. Figures 3(a)-(d) compare the results for the YIG/Nb sample (red symbols) and the YIG/AlO$_x$/Nb sample (blue symbols) at different temperatures. During all measurements, the superconducting state of the Nb modulator was continuously monitored using a nanovoltmeter to confirm that the applied supercurrent remained below the superconducting critical current. The results presented here were obtained while the Nb modulator was in the superconducting state. In the YIG/Nb sample, $A_{2\omega}$ decreases monotonically with increasing supercurrent at 2 K and 3 K. In contrast, $A_{2\omega}$ remains unchanged as a function of the supercurrent at 4 K and 5 K. The absence of modulation at 4 K and 5 K is discussed later. Moreover, no supercurrent dependence is observed at any temperature in the YIG/AlO$_x$/Nb control sample. This result rules out Joule heating of the contact electrodes and indicates that the modulation originates from an interfacial interaction between magnons and the superconductor. A possible extrinsic origin of the observed magnon modulation is the Oersted magnetic field generated by the supercurrent applied to the Nb wire. However, this possibility is unlikely because a clear modulation of the magnon signal is observed only in the YIG/Nb sample, whereas no such modulation is observed in the YIG/AlO$_x$/Nb control sample, in which the direct interaction between Nb and YIG is suppressed by the inserted AlO$_x$ spacer while a similar Oersted field is generated. Our calculations show that the Oersted fields in the two samples differ only slightly and are more than two orders of magnitude smaller than the applied external magnetic field (30 mT) used in the magnon transport measurements. Therefore, the Oersted-field contribution is negligible and cannot account for the observed modulation of the magnon signal in the YIG/Nb sample (see Appendix C).

## V. THEORETICAL MODEL

To quantitatively examine whether the modulation of magnon transport observed in Fig. 3 can be attributed to the suppression of the superconducting gap, we consider a model on the supercurrent $I_s$, as illustrated in Fig. 4(a). The quantities $A(I_s)$ and $B$, obtained experimentally, correspond to the Pt-detector voltage measured when the coupling of YIG to the superconductor is present (i.e., YIG/Nb) under an applied supercurrent $I_s$, and when the coupling is negligible (YIG/AlO$_x$/Nb), respectively. In addition, we introduce the quasiparticle spin current density generated in the superconductor under an applied supercurrent, $J_s^Q(I_s)$, which is calculated theoretically to describe the dissipation of magnon into the superconducting layer. We obtained $J_s^Q(I_s) \propto B - A(I_s)$, assuming that all of the conversion efficiency between magnons and quasiparticle spin current density in the superconductor, the magnon attenuation in YIG, and the conversion from magnons to voltage at the Pt detector interface are constant. Possible corrections to $B$, including contributions from other conversion factors in Pt, are discussed in Appendix D. To compare the experimental results with the theoretical model without relying on the unknown conversion efficiency, we define the normalized modulation amplitude $M(I_\mathrm{s})$. Since $B - A(I_\mathrm{s})$ is proportional to $J_\mathrm{s}^\mathrm{Q}(I_\mathrm{s})$, taking the ratio to the zero-current value gives

$$M(I_\mathrm{s}) \equiv \frac{J_\mathrm{s}^\mathrm{Q}(I_\mathrm{s})}{J_\mathrm{s}^\mathrm{Q}(0)} = \frac{B - A(I_\mathrm{s})}{B - A(0)}. \quad (1)$$

*shiota-y@scl.kyoto-u.ac.jp
†ono@scl.kyoto-u.ac.jp

The reduction of the superconducting gap under a finite $I_\mathrm{s}$ may be estimated as follows [29]:

$$\Delta(T, I_\mathrm{s}) = \Delta_0(T)\sqrt{1 - I_\mathrm{s}^2/I_\mathrm{c}^2}. \quad (2)$$

Here, $T$ is the temperature, $I_\mathrm{c}$ is the superconducting critical current, and $\Delta_0(T)$ is the superconducting gap at $I_\mathrm{s} = 0$, expressed as $\Delta_0(T) = 1.76k_\mathrm{B}T_\mathrm{c}\tanh\left(1.74\sqrt{1 - T/T_\mathrm{c}}\right)$ [30,31]. To incorporate the Joule heating in the Pt injector, the temperature increase was estimated, yielding temperature rises of 1.35 K at 2 K, 0.96 K at 3 K, 0.75 K at 4 K, and 0.71 K at 5 K (see Appendix E). Using the above conditions, we extend the model of the previous studies [19, 20, 32, 33] and obtain $J_\mathrm{s}^\mathrm{Q}(I_\mathrm{s})/J_\mathrm{s}^\mathrm{Q}(0)$ as follows by taking the gap suppression into account:

$$\frac{J_\mathrm{s}^\mathrm{Q}(I_\mathrm{s})}{J_\mathrm{s}^\mathrm{Q}(0)} = \frac{\int_{-\infty}^{\infty} d\epsilon \left[1 + \frac{\Delta(T, I_\mathrm{s})^2}{\epsilon(\epsilon+\delta\mu)}\right][f(\epsilon) - f(\epsilon+\delta\mu)]n(\epsilon)n(\epsilon+\delta\mu)}{\int_{-\infty}^{\infty} d\epsilon \left[1 + \frac{\Delta(T, 0)^2}{\epsilon(\epsilon+\delta\mu)}\right][f(\epsilon) - f(\epsilon+\delta\mu)]n(\epsilon)n(\epsilon+\delta\mu)}. \quad (3)$$

In Eq. (3), $\epsilon$ is the quasiparticle energy defined relative to the Fermi level, the $1 + \frac{\Delta^2}{\epsilon(\epsilon+\delta\mu)}$ is the superconducting coherence factor [33]. The normalized quasiparticle density of states is $n(\epsilon) = \frac{|\epsilon|}{\sqrt{\epsilon^2 - \Delta^2}}\theta[\epsilon^2 - \Delta^2]$, where $\theta$ is the Heaviside step function, and $f(\epsilon)$ is the Fermi distribution represented by $f(\epsilon) = \frac{1}{1+\exp[\epsilon/k_B T]}$. $\delta\mu$ is the magnon chemical potential at the interface. Figure 4(b) shows the analytical results of Eq. (3) for $\delta\mu = 0.1$, 1, 10, and 100 μeV. When determining $B$, the amplitude of thermally driven signal of local Pt was corrected (see Appendix D). The calculated results show good agreement with the experiments. The $\delta\mu$ values reported in previous studies are approximately 1–10 μeV, and the calculation results showed little variation within this realistic range of $\delta\mu$ values [34, 35]. The discrepancy between the experimental and calculated results at high current arises because Eq. (2) is valid only within a limited current range.

The absence of such modulation at 4 K and 5 K in the YIG/Nb sample can be attributed to inefficient magnon-induced quasiparticle excitation at $I_\mathrm{s} = 0$ mA, as shown in Fig. 2(b). In general, the thermally excited quasiparticle population increases with temperature. In the present model, however, the supercurrent-induced modulation is governed not simply by the total quasiparticle population, but by the change in the magnon-induced quasiparticle excitation rate, which depends on the overlap of the quasiparticle DOS, $n(\epsilon)n(\epsilon + \delta\mu)$, in Eq. (3). At 4 K and 5 K, this relevant DOS overlap is already small at $I_\mathrm{s} = 0$ mA. Consequently, even when the superconducting gap is additionally suppressed by a finite supercurrent, this overlap is not significantly enhanced, and therefore neither the magnon-induced quasiparticle excitation nor the magnon dissipation increases appreciably, leaving $A(I_\mathrm{s})$ nearly unchanged. Furthermore, no such decrease was observed in the YIG/$AlO_x$/Nb control sample at any temperature. This demonstrates that direct interfacial coupling between the YIG and the Nb plays a crucial role in the supercurrent-induced modulation of magnons. These results therefore support the interpretation that the decrease of $A_{2\omega}$ originates from enhanced magnon dissipation into the superconductor caused by the supercurrent-induced suppression of the superconducting gap.

## VI. SUMMARY

In summary, we demonstrate supercurrent-controlled magnon transport in a ferromagnet/superconductor heterostructure. The magnon transport signal decreases with increasing supercurrent. This behavior is quantitatively explained by the supercurrent-induced suppression of the superconducting energy gap, which enhances magnon dissipation into the superconductor. Our results establish experimental evidence of an interaction between magnons and superconductivity, identifying the gap energy as a tunable parameter for magnon transport. The modulation efficiency could potentially be enhanced by optimizing the superconducting layer thickness and operating temperature to increase the sensitivity of the superconducting gap to the applied supercurrent [20]. The supercurrent-controlled modulation demonstrated here may provide a route toward a superconducting gate for magnon transport and reconfigurable superconducting–magnonic circuits, in which magnon transport is controlled by a supercurrent without Joule heating. More broadly, our work introduces a new paradigm for manipulating collective spin excitations via superconducting correlations, opening a pathway toward energy-efficient magnonic and superconducting devices.

## ACKNOWLEDGMENTS

*shiota-y@scl.kyoto-u.ac.jp

†ono@scl.kyoto-u.ac.jp

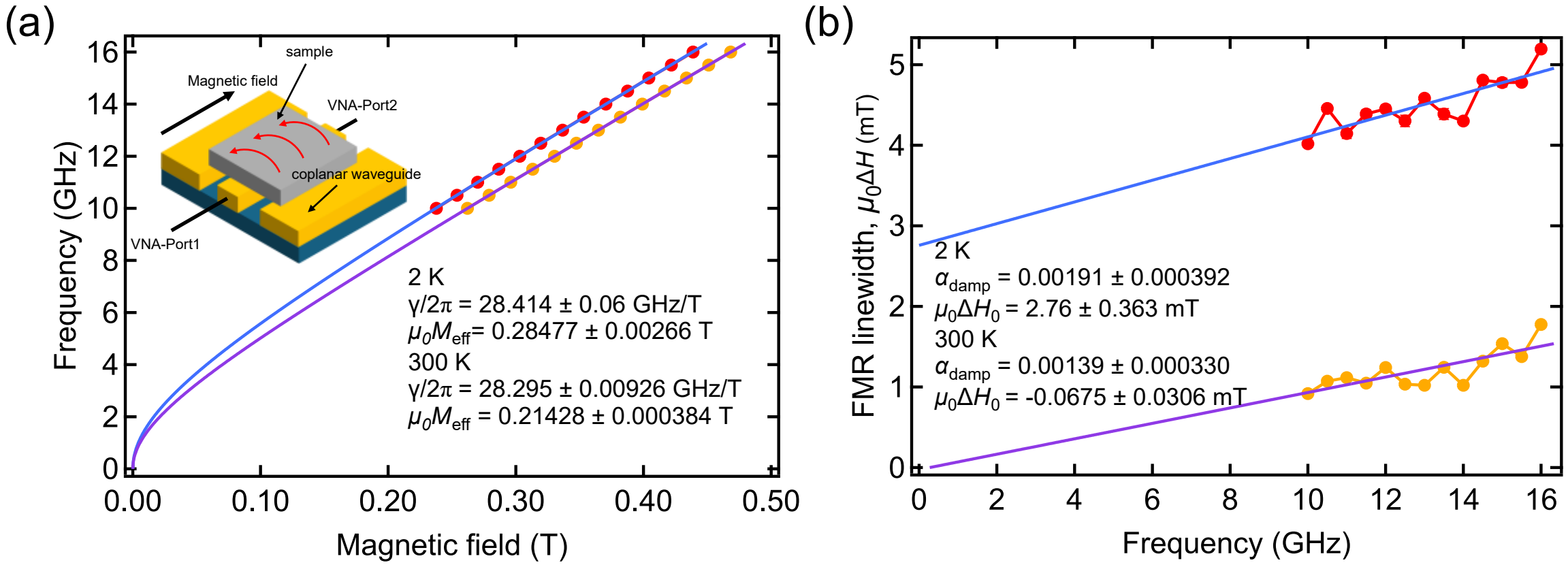

FIG. 5. (a) Resonance frequency as a function of magnetic field with the fitting results based on Kittel formula. Inset shows the schematic illustration of VNA-FMR measurement setup. (b) The full width at half maximum as a function of the microwave frequency. The solid lines represent linear fits used to extract the Gilbert damping constant $\alpha_{\mathrm{damp}}$ and the inhomogeneous linewidth $\Delta H_0$ at 2 K and 300 K. In both panels, red and orange symbols represent the data at 2 K and 300 K, respectively, while blue and purple solid lines show the corresponding fits.

This work was partially supported by JSPS KAKENHI (24H00007, 24H02233, 25K01288, 26KJ1438), MEXT Initiative to Establish Next-generation Novel Integrated Circuits Centers (X-NICS) (JPJ011438), JST SPRING (JPMJSP2110), JST PRESTO (JPMJPR25H3), and Collaborative Research Program of the Institute for Chemical Research, Kyoto University.

## DATA AVAILABILITY

There is no publicly available research data or software supporting this manuscript. Requests for further information or data should be sent to the authors.

## APPENDIX A: DAMPING CONSTANT OF YIG DETERMINED BY VNA-FMR

Figure 5(a) shows the magnetic field dependence of the resonance frequency measured by a vector network analyzer ferromagnetic resonance (VNA-FMR) at 300 K and 2 K. The measurements were performed by mounting an unprocessed YIG film on a printed circuit board. The resonance peaks in absolute value of the transmission parameter, $|S_{21}|$, were fitted using the Lorentzian function with a linear background $|S_{21}| = S\frac{(\Delta H)^2}{4(H-H_{\mathrm{res}})^2+(\Delta H)^2} + A\frac{2(H-H_{\mathrm{res}})\Delta H}{4(H-H_{\mathrm{res}})^2+(\Delta H)^2} + C_0 + C_1 H$ , where $H$ is the external magnetic field; $S$ and $A$ are the symmetric and antisymmetric components, respectively; and $H_{\mathrm{res}}$ is the resonance magnetic field; $\Delta H$ is the full width at half maximum [36]. The resonance magnetic fields were analyzed using Kittel formula $f = \frac{\mu_0\gamma}{2\pi}\sqrt{H_{\mathrm{res}}(H_{\mathrm{res}} + M_{\mathrm{eff}})}$ [37,38]. Here, $\gamma$ is the gyromagnetic ratio and $M_{\mathrm{eff}}$ is the effective magnetization. Figure 5(b) shows the full width at half maximum of the resonance peak as a function of the microwave frequency. The damping constant $\alpha_{\mathrm{damp}}$ was extracted based on the equation of $\Delta H = \Delta H_0 + \frac{4\pi\alpha_{\mathrm{damp}}}{\mu_0\gamma} f$ [38]. We find that $\alpha_{\mathrm{damp}}$ increases at 2 K compared to 300 K. This enhancement is attributed to a paramagnetism of the GGG substrate, becoming particularly pronounced at low temperatures [39].

## APPENDIX B: ESTIMATION OF COERCIVE FIELD OF YIG

Figure 6 shows the magnetic field dependence of the thermally driven voltages, $V_{\mathrm{th}} = \frac{V(+I)+V(-I)}{2}$, of the Pt wire measured when a dc current of $I = 1$ mA ($3.33 \times 10^{10}$ A/m$^2$) was applied to the Pt wire on the YIG. The magnetic field was applied along $\alpha = 0°$, perpendicular to the current direction. In this configuration, a magnetization-dependent voltage is detected due to the spin Seebeck effect [40]. From these hysteresis loops, the coercive field of the YIG was estimated to be approximately 0.2 mT at 2–5 K. Note that the residual field of the apparatus was considered, and the hysteresis loop was corrected so that its center coincides with zero field.

## APPENDIX C: ESTIMATION OF THE OERSTED MAGNETIC FIELD BY SUPERCURRENT

The YIG/Nb sample and the YIG/$AlO_x$/Nb sample differ by the insertion of a 20 nm thick $AlO_x$ layer between the YIG and Nb layers. Therefore, a possible extrinsic origin of the observed magnon modulation is the Oersted magnetic field generated by the applied supercurrent. To evaluate the influence of the Oersted field generated by the supercurrent, we estimated the Oersted magnetic field at the YIG interface for both samples [41]. The maximum amplitude of the supercurrent applied in this study was 1.5 mA, corresponding to a current density of $7.5 \times 10^9$ A/m$^2$. Figure 7 shows the spatial distribution of the Oersted magnetic field calculated for 1.5 mA. The maximum difference in the Oersted magnetic field between two samples is approximately 0.008 mT in the in-plane field ($x$) direction and 0.018 mT in the out-of-plane ($z$) direction. These values are more than three orders of magnitude smaller than the applied magnetic field of 30 mT used in the nonlocal magnon measurements. Moreover, such small field differences are insufficient to induce measurable changes in magnon excitation, propagation, or magnetic configuration of the YIG. We therefore conclude that the observed modulation of the nonlocal magnon signal cannot be attributed to the Oersted magnetic field generated by the supercurrent.

*shiota-y@scl.kyoto-u.ac.jp
†ono@scl.kyoto-u.ac.jp

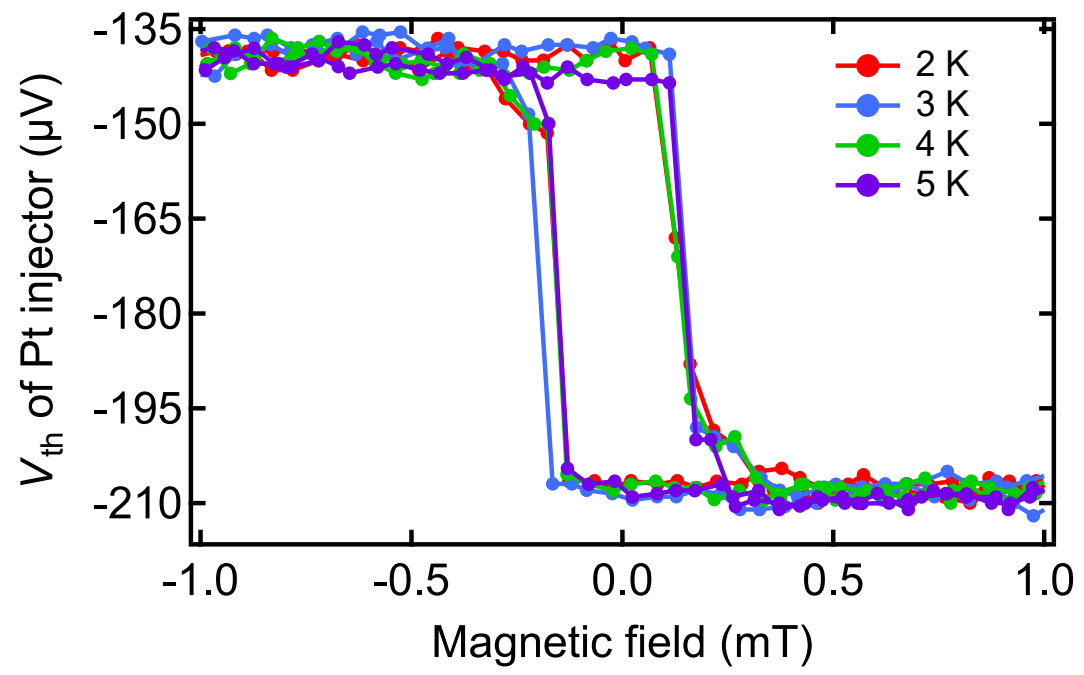

FIG. 6. Magnetic field dependence of the thermally driven voltage $V_{\text{th}}$ in the Pt wire of the YIG/Nb sample at 2-5 K.

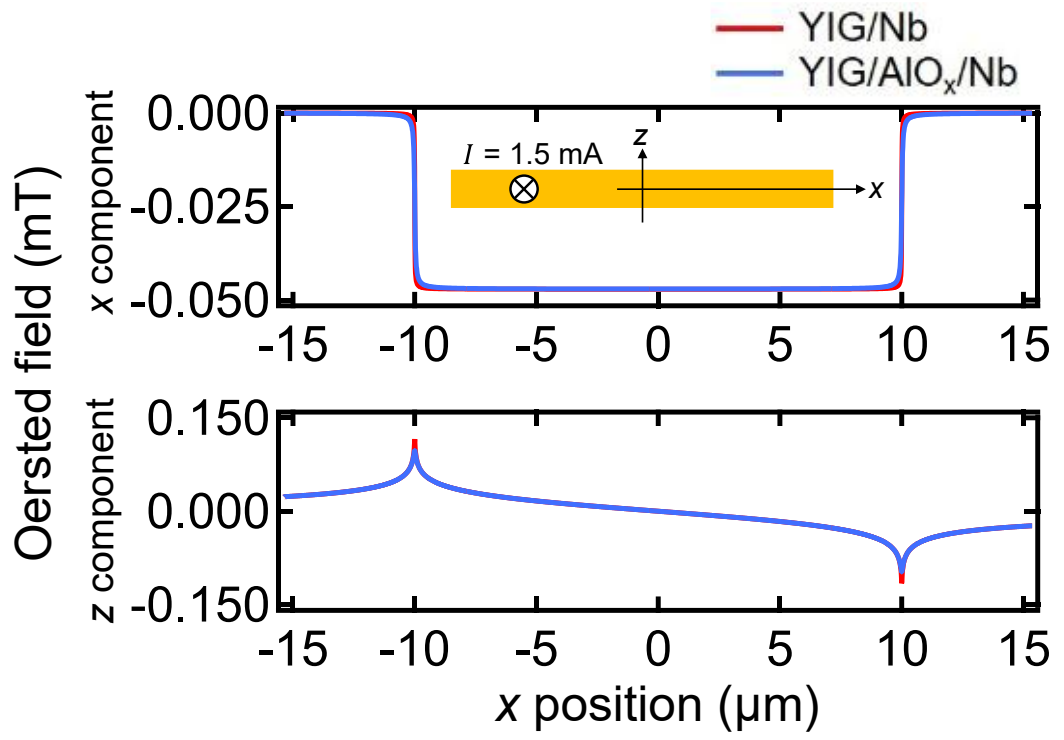

FIG. 7. Calculated spatial distribution of the Oersted magnetic field generated by a supercurrent of 1.5 mA ($7.5 \times 10^9$ A/m$^2$).

## APPENDIX D: CORRECTION OF LOCAL PT CONDITIONS ACROSS DIFFERENT DEVICES

The relation $J_s^Q(I_s) \propto B - A(I_s)$ assumes only that the change in the detected signal is proportional to the quasiparticle spin current. In practice, however, the experimentally obtained value of $B$ must be corrected for device-to-device variation of the resistance, spin Hall angle, and interfacial spin-mixing conductance even when Pt wires are fabricated on the same YIG film. Therefore, when comparing $A(I_s)$ and $B$ to obtain $\frac{B-A(I_s)}{B-A(0)}$, correction is required. In this study, we consider only the thermally driven effect. Therefore, we account for device-to-device variations and compare the local amplitude of the α dependence derived from the local spin Seebeck effect between the two devices. Figure 8 shows the magnetic field angle dependence of $V_{\text{th}}$ of the local Pt in the device used in the experiment with and without Nb used to determine $B$ under $\mu_0 H_{\text{ext}} = 30$ mT at 2 K. The amplitudes for the device with and without Nb were 40.584 μV and 32.186 μV, respectively. Since the contribution of Joule heating appears as $VI = RI^2$ and is proportional to $I^2$, the correction factor for the applied current value is given as $\sqrt{\frac{40.584}{32.186}}$. Accordingly, when determining $B$, a current of 1.347 mA ($4.49 \times 10^{10}$ A/m$^2$), obtained by multiplying the current value by this factor, was applied.

## APPENDIX E: ESTIMATION OF TEMPERATURE RISE INDUCED BY AC CURRENT APPLIED TO THE PT INJECTOR

Figure 9 shows the superconducting critical current $I_c$ of Nb wire with and without ac current applied to Pt, along with the fitting results for the case without applied current. For the fitting, the following equation was used [29]:

$$I_c(T) = I_c(0) \times \left[1 - \left(\frac{T}{T_c}\right)^2\right]^\beta \qquad \text{(E1)}$$

The experimentally determined $T_c$ was 6.5 K, and $\beta = 0.504 \pm 0.012$ was obtained from the fitting. The temperature rise caused by applying ac current to Pt was estimated by comparing the measured $I_c$ under ac bias current with the fitted $I_c(T)$ curve and determined to be 1.35 K at 2 K, 0.96 K at 3 K, 0.75 K at 4 K, and 0.71 K at 5 K based on the fitting results.

*shiota-y@scl.kyoto-u.ac.jp
†ono@scl.kyoto-u.ac.jp

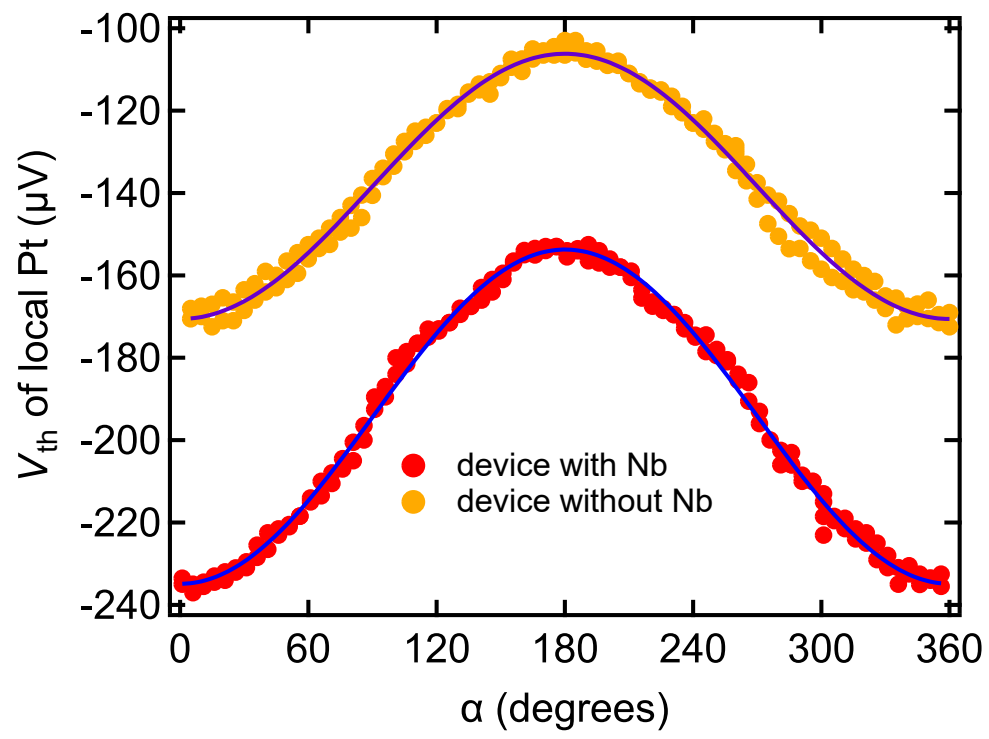

FIG. 8. Magnetic field angle dependence of $V_{\mathrm{th}}$ of local Pt for the devices with and without Nb under $\mu_0 H_{\mathrm{ext}} = 30$ mT at 2 K. Red dots indicate the device with Nb, while orange dots indicate the device without Nb. The solid blue and purple lines represent the fitting results, respectively.

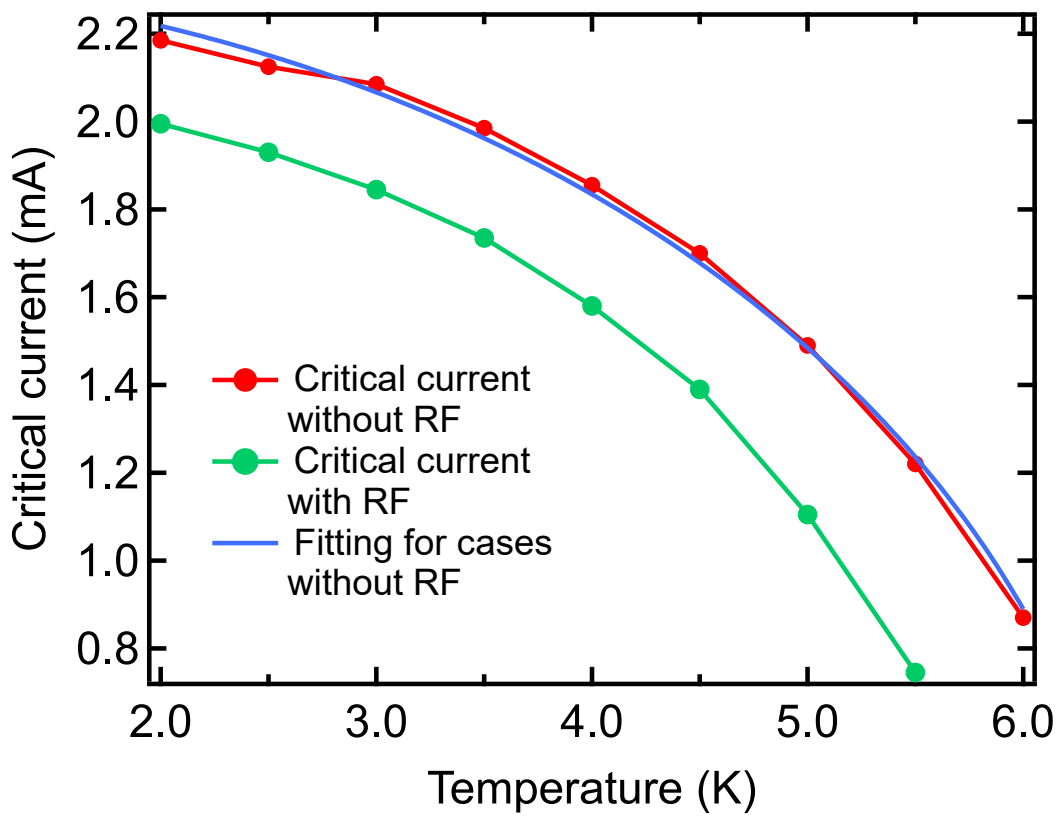

FIG. 9. Change in the superconducting critical current due to Joule heating induced by the ac current applied to the Pt injector. The red dots indicate the critical current without applied current, while the green dots indicate the critical current with ac current applied. The blue solid line represents the fitting result using Eq. (E1)

*shiota-y@scl.kyoto-u.ac.jp

†ono@scl.kyoto-u.ac.jp

*shiota-y@scl.kyoto-u.ac.jp
†ono@scl.kyoto-u.ac.jp

*shiota-y@scl.kyoto-u.ac.jp
†ono@scl.kyoto-u.ac.jp